%% file: amspaperV6.1.tex
\documentclass{ametsocV6.1}

\title{Quantifying the Effect of a Parallax Correcting Algorithm for Passive Microwave Satellite Precipitation Retrievals across the Continental United States}

\authors{Andres F. Monsalve\aff{a}, Hernan A. Moreno\aff{a}\correspondingauthor{Hernan A. Moreno, moreno@utep.edu}, Eric Goldenstern\aff{b}, Christian Kummerow\aff{b} 
}

\affiliation{\aff{a}{Department of Earth, Environmental and Resource Sciences, University of Texas at El Paso}\\
\aff{b}{Department of Atmospheric Science, Colorado State University}
}

\abstract{Satellite precipitation retrieval algorithms whose measurement instruments are tilted to the zenith line are subject to a spatial mismatch between the theoretical ground coordinates and the coordinate pair corresponding to the cloud layers sending spectral signals to the satellite. This is the case of the precipitation retrievals of the GPM Passive Microwave Imagery (GMI) on board the core satellite of the Global Precipitation Mission (GPM) that uses the Goddard Profiling Algorithm (GPROF). Currently, no geometrical correction is applied to GMI retrievals of surface precipitation, creating a horizontal displacement (or parallax mismatching) between the reported surface and the corrected coordinates corresponding to the cloud structures intersecting the field of view. 
 GPROF precipitation retrievals over the Continental United States are analyzed using the ground-validated Multi-Resolution Multi-Sensor (GV-MRMS) system data and the European Centre for Medium-Range Weather Forecasts Reanalysis version 5 (ERA5) temperature profiles. Results applying this parallax correction scheme show improvements in the overall retrieval accuracy of GPROF, mainly during the summer months, for every precipitation type, when the freezing level (FL) is relatively high. The development of this new parallax-correction algorithm for passive microwave radiometers will significantly improve the accuracy of remote sensing data by minimizing spatial distortions in atmospheric measurements, leading to more precise weather forecasting, climate monitoring, and environmental assessments.} 

\usepackage{hyperref}
\usepackage{etoolbox} 
\newrobustcmd{\internallinenumbers}{}
\newrobustcmd{\linenomath}{}

\begin{document}

\maketitle

%
\statement
Our study addresses a critical issue in satellite precipitation retrieval algorithms regarding the spatial mismatch from tilted measurement instruments. By examining the precipitation retrievals of the Global Precipitation Mission Microwave Imager (GMI) instrument, we highlight the consequential parallax mismatching between reported surface precipitation from observed cloud structures, as well as the resulting error from this discrepancy. Through a novel geometrical correction scheme applied to GMI retrievals over the Continental United States, we demonstrate significant enhancements in retrieval accuracy, particularly during summer months that are characterized by high-altitude freezing levels. These findings advance the reliability of satellite-based precipitation estimation, which is crucial for meteorological research and operational forecasting.

%

\input{1_introduction}

\input{2_methods}

\input{3_results}              
\input{4_discussion}

\input{5_conclussions}

\clearpage

\acknowledgments
\

This research has been supported by the National Aeronautics and Space Administration (NASA), Precipitation Measurement Mission program (grant no. 80NSSC22K0604).


\datastatement
\

The Python implementation of the height-based parallax correction algorithm based on the freezing level is available at \textit{https://github.com/afmonsalves/parallax\_correction\_GMI.git}. GPROF level 2A retrieval results data are available at https: \textit{//doi.org/10.5067/GPM/GMI/GPROFCLIM/2A/07} \citep{GPROF_data}. The ground-validated MRMS dataset over the CONUS are available at \textit{https://pmm-gv.gsfc.nasa.gov/} \citep{GVMRMS_data}, and the ERA5 data was retrieved from \textit{https://cds.climate.copernicus.eu/cdsapp\#!/dataset/reanalysis-era5-pressure-levels} \citep{Hersbach_ERA5}.



\bibliographystyle{references_style}
\bibliography{references}

\end{document}

%% file: 1_introduction.tex
\section{Introduction}

Satellite retrievals have been used in the last three decades as a viable method to characterize the spatial distribution of precipitation across the globe. In the mid-1980s, a series of multifrequency microwave sensors were launched to provide continuous precipitation measurements (Kidd and Levizzani, 2011). These have been enhanced by dedicated precipitation missions such as the Tropical Rainfall Measuring Mission (TRMM) launched in 1997, and the Global Precipitation Measurement (GPM) mission, launched in 2014. Both these missions carried precipitation radars in addition to the passive microwave sensors that provide continuous precipitation estimates based on measured Brightness Temperature (TB).

The GPM Core Observatory Dual-frequency Precipitation Radar (DPR), developed by JAXA, and the GPM Microwave Imager (GMI), developed by NASA and Ball Aerospace, work together to provide comprehensive precipitation observations from above the clouds. The DPR aboard the GPM Core Observatory uses radar pulses to measure the three-dimensional structure of precipitation, including its intensity and vertical distribution. DPR provides detailed information about the size and shape of raindrops, snowflakes, and ice particles within a storm system. The GMI instrument complements the DPR by providing passive microwave measurements of precipitation. GMI operates at 13 different frequencies, allowing for the estimation of precipitation rates and the detection of frozen and liquid precipitation particles \citep{kojima_dual-frequency_2012,bidwell_global_2005}. In addition to the GPM core observatory, the GPM mission includes a network of international partner satellites; the partner satellites contribute additional microwave and infrared precipitation measurements, combined with the GPM core observatory data, to create a unified global precipitation dataset consisting of two merged radar-passive microwave conical scan imager products, six conical-scan passive microwave imagers, six cross-track-scan passive microwave sounders, five geosynchronous infrared imagers, three infrared/passive microwave sounders, and monthly accumulated rain gauge information where possible. \citep{IMERG_doc}.

The Goddard profiling (GPROF) is the standard algorithm for all the passive microwave (PMW) sensors of the GPM framework. It is a physically based Bayesian precipitation retrieval algorithm that uses a prior database of cloud and precipitation profiles from radar-derived cloud structures observed by TRMM and GPM \citep{nasa_gpm_2022}.

\nocite{kummerow_evolution_2015}
\nocite{guilloteau_global_2017}
\nocite{pfreundschuh_gprof_2024}

\begin{figure}[h!]
  \centering
  \includegraphics[width=0.7\textwidth]{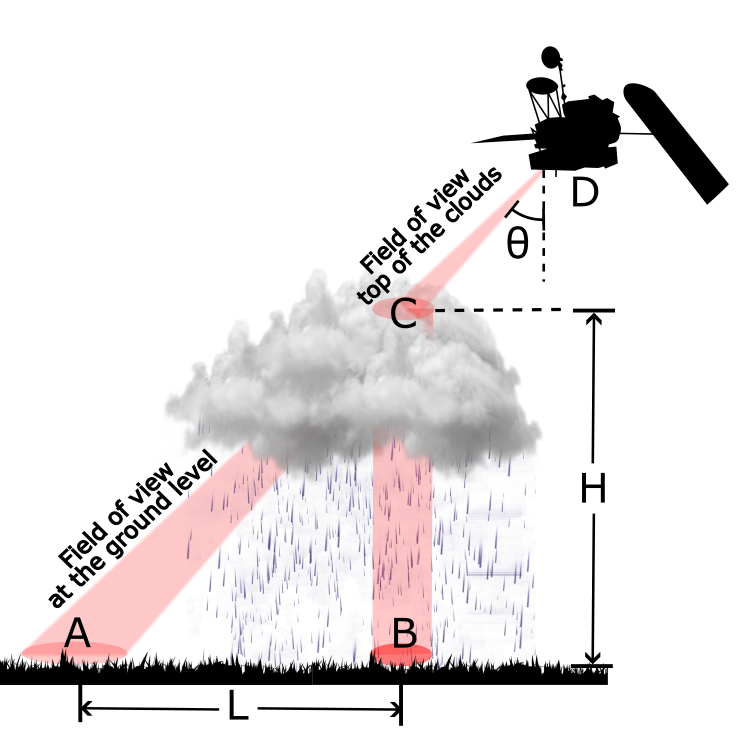}
  \caption{Parallax mismatching caused by the off-nadir field of view intersection with the cloud structure. D represents the vertical height of the satellite above the earth's ground surface, $\theta$ the angle of the satellite field of view with the vertical (i.e., nadir line), C the cloud-top first intersected by the satellite field of view, A the corresponding ground coordinates of the retrieved signal, B the observed precipitation corresponding to the cloud-top directly (i.e., vertically) below C, H represents the height of the cloud intercepting the FOV of the satellite, and L the distance between A and B.}
  \label{fig:parallax_sketch}
\end{figure}

Numerous studies, such as those by Kummerow et al. (2015), Guilloteau et al. (2017), and Pfreundschuh et al. (2024) have validated GPROF results in various regions worldwide. However, a universal source of uncertainty for GMI and other scanners is the parallax shifting caused by the instrument's conical field-of-view (FOV) and its intersection with the clouds. The parallax shifting causes a horizontal displacement between the apparent coordinates of the retrieval, calculated as the intersection of the FOV with a geoid-height model, and the actual coordinates of the observed precipitation by radars or rain gauges \citep{hughes_comparison_2006}. This problem is summarized in Figure \ref{fig:parallax_sketch}. 

Consider a satellite with a field of view pointing off-nadir with an angle $\theta$. The first object that the satellite comes across is the top of any cloud at point \textbf{C}; the corresponding measured TB is related by the algorithm to the point \textbf{A}, projected into the ground following the original trajectory of the field of view. Therefore, the problem is that the assigned coordinates (i.e., those of point \textbf{A}) associated with the TB measured at \textbf{C} are representative of the precipitation near \textbf{B}. The horizontal distance between \textbf{A} and \textbf{B} is defined as the parallax mismatch. The parallax mismatch increases as a function of the height of the point \textbf{C}, or the cloud's intercept height.

Both geostationary and polar-orbiting satellites are susceptible to this kind of error as most of them do not account for the height at which the measured signal originates \citep{masunaga_inter-product_2019}. In infrared methods, a common approach is to perform a parallax correction (PC) based on the measured cloud temperature and derivation of the correction height based on a hypothetical vertical distribution of atmospheric temperature, pressure, and density known as the Standard U.S. Atmosphere \citep{vicente_role_2002, eumetsat_koenning}. An alternate approach is to use ancillary information as a nadir-oriented auxiliary satellite to detect the top of the cloud and then collocate it with the satellite of interest as with the collocation of MODIS data and CloudSAT data in Wang et al. (2011). Cloud-top-based corrections and partner satellite collocation approaches have shown usefulness in reducing the measured errors of the satellite retrievals and are being used regularly by multiple agencies to correct their retrievals \citep{satpy_python_library}. However, the methodologies above do not involve the physical principle of functioning of PMW instruments that account for the ice layer inside the cloud structures. A cloud-top approach may be suitable for infrared sensors but is not optimal for every sensor, as is the case for PMW.
\nocite{wang_parallax_2011}

This research aims to look into the PC for microwave radiometers, further understand the magnitude of the parallax mismatching, its seasonality, and regional distribution within the U.S., and the improvement of the corrections when a physically-based algorithm is applied. The proposed methodology utilizes the ice layer in the atmospheric profile, specifically within the cloud. The ice layer is responsible for the maximum brightness temperature depression for a PMW sensor and should account for the best possible PC. The PC algorithm's seasonality and regime dependency will be investigated, and its usefulness will be addressed.

This study describes a parallax-correcting algorithm with a systematic quantification of improvements via validating satellite microwave-derived precipitation in different cloud regimes (i.e., convective, stratiform, snow, and hail). Native pixel-level resolution in the Continental United States (CONUS) data from a full year of GMI instrument using GPROF is used to assess the improvements after including the correcting scheme. In addition to cloud types and heights, the algorithm accounts for the geometry of the FOV in relation to the footprint of the benchmark data. Section 2 of this article presents the used data and methodological steps of the correcting algorithm, while Sections 3, 4, and 5 present the results, discussion, and conclusions of this research.

%% file: 2_methods.tex
\section{Data and Methods}

\subsection*{Study Region} 

The proposed algorithm was tested over the whole CONUS and over five 10°x10° sub-regions to isolate cloud-specific regional impacts. The regions selected to represent diverse precipitation regimes across the US were: Northwest-NW, Southwest-SW, Central Plains-CP, Southeast-SE, and Northeast-NE (Figure 2).

\begin{figure}[h]
  \centering
  \includegraphics[width=1\textwidth]{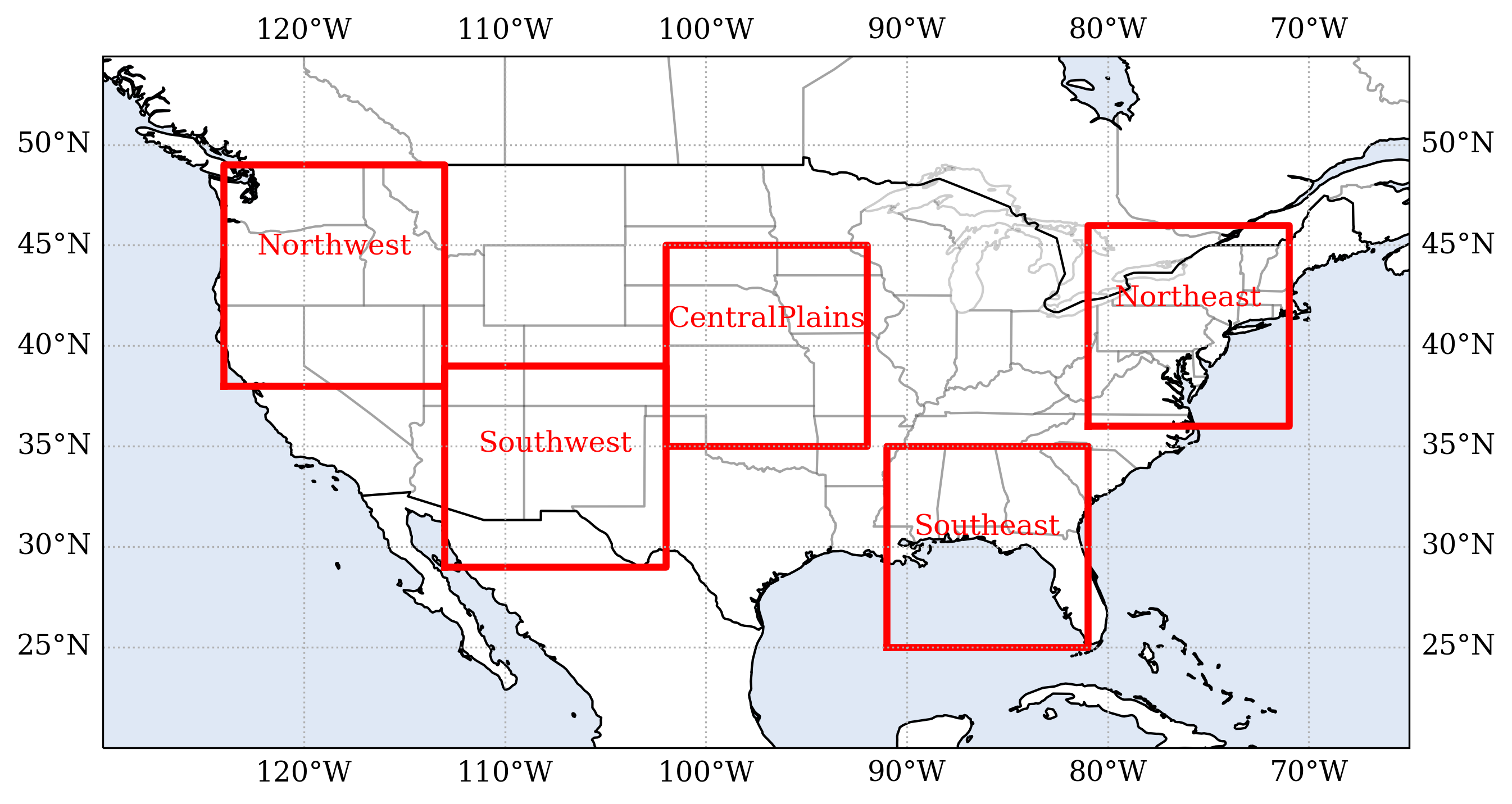}
  \caption{Bounding boxes for the five study regions inside the CONUS. }
  \label{fig:figure_location}
\end{figure}

\subsection{Data} 

Three types of data are used to conduct the primary analyses: (1) precipitation retrievals from GPM, (2) the Ground Validated Multi-Radar Multi-Sensor precipitation retrievals (GV-MRMS) benchmark product, and (3) a set of multi-level environmental weather variables from reanalysis outputs. An additional data source is included to perform a comparative analysis with another method of PC; (4) Cloud-top height data from the Geostationary Operational Environmental Satellite - R series (GOES 16).

\subsubsection{Precipitation Retrievals from GPM}

Precipitation rates from GMI were retrieved using GPROF version 7 \citep{nasa_gpm_2022} between 10/01/2019 and 09/30/2020. The satellite carrying the GMI instrument has a non-sun synchronous 65° inclination orbit and completes an entire revolution around the Earth in roughly 90 minutes. The antenna is pointed at 45° from the vertical, resulting in an Earth incidence angle of approximately 53°. The GMI instrument rotates around its vertical axis and completes one revolution every two seconds, collecting measurements over roughly 1/3 of the revolution with a scan width of 885 km. The instrument has 13 passive microwave channels ranging in frequency from 10.65 GHz to 183 GHz, allowing it to detect mixed types of precipitation with intensities ranging from 0.2 mm/h to 110 mm/h  \citep{skofronickjackson_global_2018}. The satellite constellation that conforms GPM has an average revisit time of three hours over 80\% of the earth's surface. For additional details on GMI functioning, see NASA (2022).

\nocite{nasa_gpm_2022}
\subsubsection{Benchmark Multi-Sensor Precipitation Retrievals}

The GV-MRMS quantitative precipitation dataset was used as a benchmark product for comparison and evaluation of the proposed PC. This product contains multiple datasets that are used in the proposed approach: (1) precipitation rate, (2) type estimates, and (3) quality control. The GV-MRMS incorporates data from the weather surveillance radar 88 Doppler, the rapid update cycle model analysis fields, and rain gauge data, as does the original Multi-Radar Multi-Sensor (MRMS), with an additional hourly rain-gauge-bias adjustment. The product has a spatial resolution of 1 km x 1 km and a time resolution of two minutes and is maintained by the NASA Global Hydrometeorology Resource Center DAAC \citep{kirstetter__pierre-emmanuel_gpm_2018}. 

\subsubsection{Multi-level Environmental Weather Variables}

The European Centre for Medium-Range Weather Forecasts Reanalysis version 5 (ERA5) is used in this study. From ERA5, two variables were used to calculate the height of the freezing level (FL): (1) the temperature at 37 pressure levels and (2) the geopotential height at each of those levels. ERA5 data is appropriated for this purpose as its accuracy regarding temperature has been proven acceptable for hydrological purposes \citep{tarek_evaluation_2020}. It is the most reliable regular-high-resolution data source on a continental scale. ERA5 has a native resolution of approximately 25 km spatially and one hour temporally. Some of ERA5's limitations include non-physical trends and variability due to changes in the observing systems where the observations do not dominate the data assimilation \citep{Hersbach_ERA5}. These subtle trends, however, are not believed to impact the findings of this study.

\subsubsection{Cloud-Top Height from GOES 16}

GOES-16 cloud-top height data with a resolution of 10km x 10km and 10 minutes is used to compare a cloud-top-based PC against a FL-based one. The dataset is derived from the satellite’s Advanced Baseline Imager (ABI), which provides high-resolution, multi-spectral imagery across visible, infrared, and near-infrared channels. The Python library Goes-2-go was used to access the data for July 2020 over the CONUS \citep{goes2go}. Additional details for the GOES dataset are available at \citep{goes-r_algorithm_working_group_noaa_2018}

\subsection{Methods}

The match-up between GV-MRMS, GPROF's FOV, and ERA5 has additional steps to ensure that their respective spatial scale, time scale, and field of view are fairly matched and compared. Those steps are outlined below:

\begin{enumerate} 
\item \textbf{Locate the FL}:
To locate the FL, the pressure at which the temperature profiles switch to 0° Celsius was converted to height using the geopotential height data from ERA5. In cold environments, the FL could be calculated at, or below, the surface level. This study did not consider pixels with an estimated FL below surface level.

\item \textbf{Weighted averaging of GV-MRMS precipitation estimates:}
To match the GMI instantaneous retrievals within its FOV with the GV-MRMS dataset (that has considerably finer spatial resolution than the GMI retrievals), we implemented a weighted averaging technique for the GV-MRMS precipitation rate data that accounted for neighboring pixels. Specifically, we assigned weights following a Gaussian distribution to each GV-MRMS pixel, centered in the pixel closest to the matched GPROF pixel with weights decreasing radially according to the theoretical GMI antenna gain function \citep{guilloteau_beyond_2020}. For temporal matching, each GMI pixel time was matched with the closest available GV-MRMS file, ensuring the time gap between the GV-MRMS file and the GMI pixel was less than one minute. If the closest GV-MRMS file was more than one minute away (i.e. half of its time resolution), the pixel was not included in the analysis. As a result, one orbit of the satellite was likely matched with multiple GV-MRMS files.

\item \textbf{Pre-processing and quality control of data:}

The analysis did not include pixels with precipitation intensities below 0.2 mm/h simultaneously for GMI and GV-MRMS due to the tendency of GMI to get noise signals falsely interpreted as light rain; 0.22 mm/h is the minimum estimated detectable precipitation intensity for any surface type for GMI \citep{munchak_evaluation_2013}. 

\item \textbf{Coordinates calculations:}

The distance between \textbf{A} and \textbf{B} in Figure \ref{fig:parallax_sketch} represents the parallax displacement. The coordinates of the point \textbf{A} are known. We used the Great Circle Distance method (GCD) to find the coordinates of the parallax-corrected precipitation retrievals at point \textbf{B}. The distance \textit{L} in Figure \ref{fig:parallax_sketch} is given by equation \ref{eqdist}, where \textit{H} is the height of the interception between the satellite FOV and the observed cloud layer and $\theta$ is the satellite FOV inclination with respect to the zenith.

\begin{equation}\label{eqdist}
    L = \frac{H}{\tan{(90 - \theta)}}
\end{equation}

The corrected latitude (\textit{\textbf{$Lat_{B}$}}) and longitude (\textit{\textbf{$Lon_{B}$}}) are calculated with GCD equations \ref{eqlat} and \ref{eqlon}  \citep{carter2002great}. An approximate Earth radius (\textit{R}) of 6371km was considered. The bearing, (\textit{b}), is the satellite FOV angle from the north in a clockwise direction.

\begin{equation}\label{eqlat}
    {Lat}_{B} = \arcsin\left(\sin ({Lat}_{A}) \cdot \cos \left(\frac{L}{R}\right)+\cos ({Lat}_{A}) \cdot \sin \left(\frac{L}{R}\right) \cdot \cos (b)\right)
\end{equation}

\begin{equation}\label{eqlon}
    {Lon}_{B} = {Lon}_{A} +\operatorname{atan}2\left(\sin (b) \cdot \sin \left(\frac{L}{R}\right) \cdot \cos ({Lat}_{A}),  \cos\left(\frac{L}{R}\right) - \sin({Lat}_{A}) \cdot \sin({Lat}_{B})\right)
\end{equation}

\item \textbf{Parallax correction heights:}

The GMI instrument measures the radiance at the top of the atmosphere resulting from the surface emission, the water vapor emission, as well as clouds and precipitation, and the subsequent scattering caused by the presence of ice hydrometeors \citep{guilloteau_beyond_2020}. To determine the height of the ice feature responsible for the apparent scattering signal, we consider the freezing level and the surrounding heights, three km above and below the freezing level (with intervals of 500 m), for a total of 13 possible best parallax correction heights. At any time in the trajectory of the GPM core observatory, GMI position, satellite height, and scanning direction are known. That information is used along with the different test heights to obtain the parallax-corrected position for each GMI pixel retrieval.

\item \textbf{GPROF improvement assessment:}

To quantify the overall improvement of the proposed parallax-correcting algorithm, two evaluations are made: (1) Identifying the altitude-correction level that yields the greatest positive improvement, along with quantifying the gains and losses across various cloud-altitude levels, and (2), a quantification of the gains and losses in estimation accuracy for different precipitation types and regimes after applying the PC found in (1). All evaluations are made in terms of the root mean squared error (RMSE) and Pearson correlation coefficient ($\rho$) on a monthly or seasonal basis with respect to GV-MRMS. 

To further assess the accuracy of the parallax-corrected GPROF outputs at the event-scale, multiple events were analyzed before and after PC and its Lacunarity calculated to quantify the heterogeneity and unevenness of the precipitation systems. Lacunarity is a measure of translational invariance developed for fractal sets, described first by Mandelbrot (1994), but it has also been used as a multi-scale description of spatial dispersion in diverse applications \citep{plotnick_lacunarity_1996}. To compute Lacunarity over a precipitation scene, a square window of size (\textit{k}) is defined and glided over the domain, finding the sum of values inside the moving window at each iteration. Later, for the resulting array of sums, their variance (\textit{var}) and mean (\textit{m}) are calculated, and using equation \ref{eqlac}, the Lacunarity value for the scale (\textit{k}) is obtained. For additional details on how to calculate Lacunarity, see Plotnick et al. (1996).

\nocite{plotnick_lacunarity_1996}
\nocite{mendelbrot_lacunarity_1994}

\begin{equation}\label{eqlac}
    {Lacunarity}_{k} = \frac{var}{m^2 + 1}
\end{equation}

\end{enumerate}

%% file: 3_results.tex
\section{Results}

\subsection{Determination of the FL}

\begin{figure}[h]
  \centering
  \hspace{3cm}
  \centerline{\includegraphics[width=0.9\textwidth]{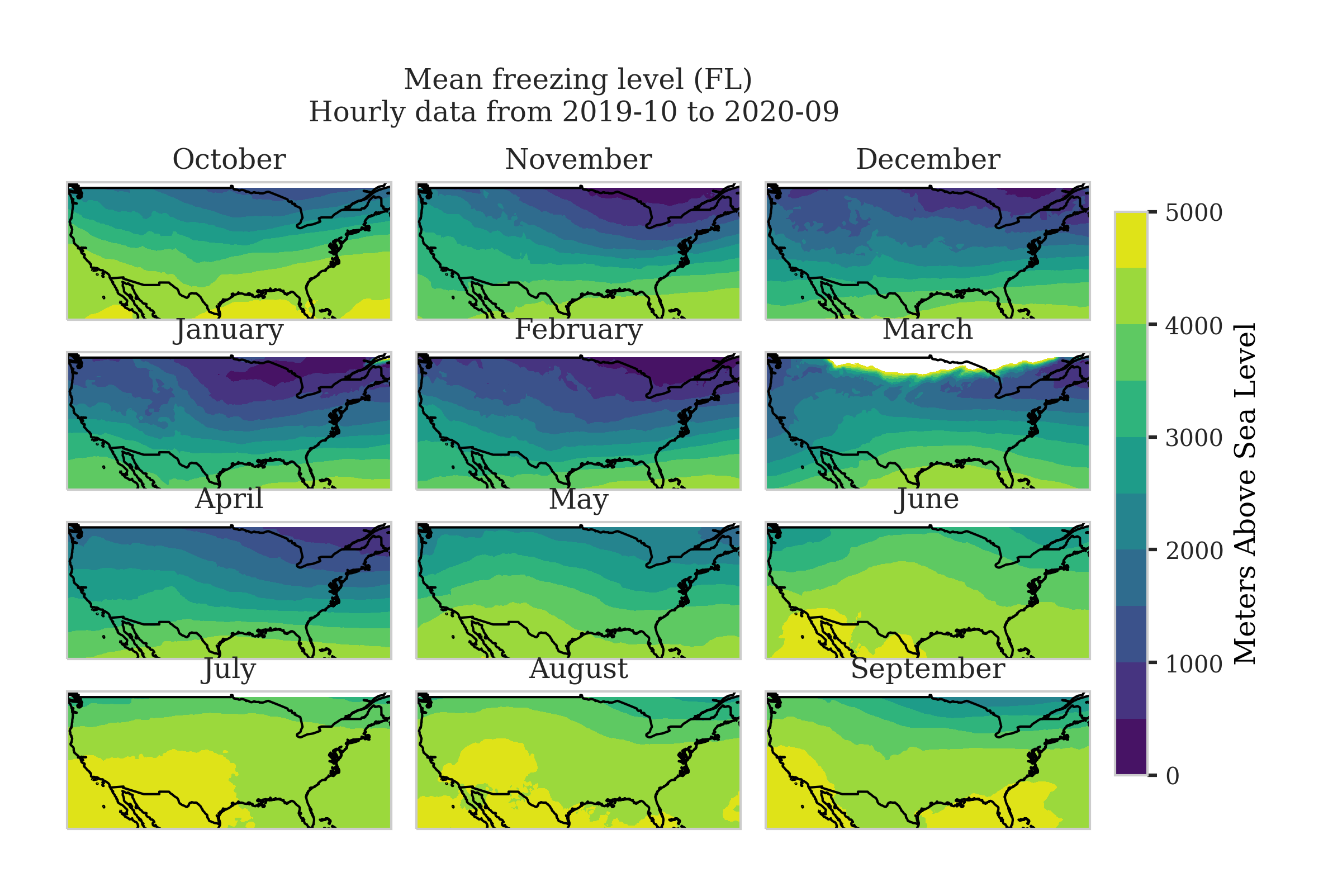}}
  \caption{Mean monthly freezing level altitude over the CONUS as computed from hourly ERA-5 temperature profiles from October 2019 to September 2020. Uncolored areas indicate that a thermal inversion was not found in the used dataset}
  \hspace{3cm}
  \label{fig:FL}
\end{figure}

Figure \ref{fig:FL} illustrates the calculated CONUS monthly average FL from October 2019 to September 2020. Throughout the year, the FL varies with mean surface temperatures. It is higher in the months following the summer solstice and lower in the months following the winter solstice. The lowest altitude of the FL (between 0 and 500 meters above sea level) occurs in January and February over the northeast section of the U.S., and the highest values  (between 4,000 and 4,500+ meters above sea level) occur from June through September across the southernmost latitudes. 

\subsection{Height-Based Parallax Correction and Quantification of Estimation Gains}

\begin{figure}
    \centering
    \includegraphics[width=0.95\linewidth]{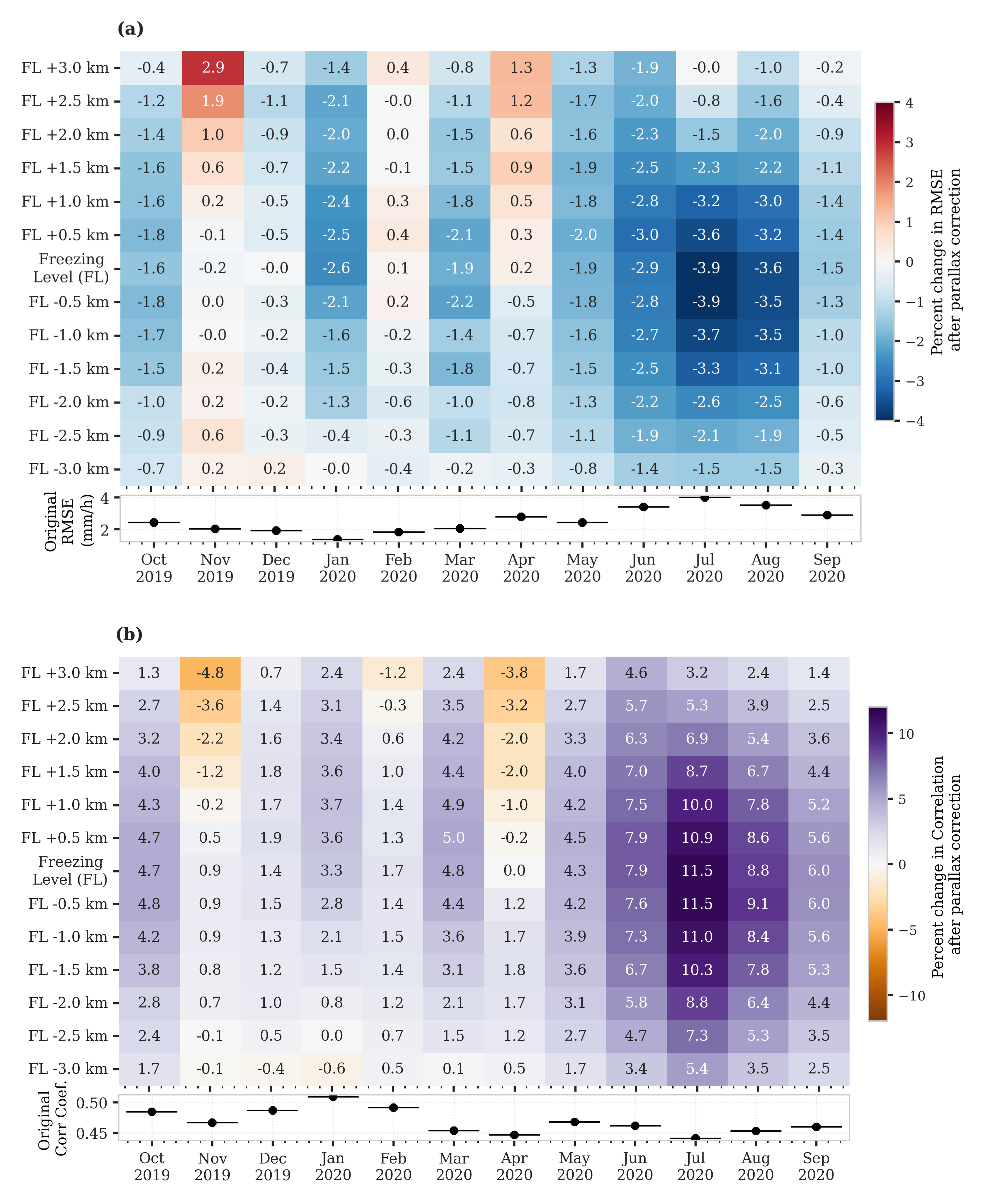}
    \caption{Mean percent change of (a) RMSE and (b) Pearson correlation coefficient ($\rho$), between GPROF and GV-MRMS precipitation intensities (mm/h) after (in relation to prior-to) applying PC with different heights to GPROF over the CONUS. Statistics were calculated and averaged at pixels where precipitation was captured either by GPROF or GV-MRMS from 10/01/2019 to 09/30/2020. The subplots below each main panel illustrate the original, pre-parallax correction RMSE and $\rho$ values, respectively.}
    \label{fig:statistics_CONUS}
\end{figure}

Thirteen different PC heights were applied across CONUS, ranging from three kilometers below the FL to three kilometers above the FL. Results, shown in Figure \ref{fig:statistics_CONUS} summarize on a monthly basis the convenience of each of the thirteen heights as a PC altitude in terms of their average RMSE and correlation.

Figure \ref{fig:statistics_CONUS}(a) shows that overall, negative changes or reductions in RMSE (blue colors) tend to occur across the CONUS around the FL, but optimal values occur at the FL, where  improvements in the estimation of precipitation are achieved after applying the PC. Without any PC, the monthly average values of RMSE (shown in the sub-panels below each main panel in figure \ref{fig:statistics_CONUS}) vary between 1.3 mm/h in January to 4 mm/h in July, with higher errors corresponding to the months with tallest freezing levels (i.e. June through September). After applying the PC at the FL, such RMSE values are reduced by 1.6 \% on average over the CONUS, with a maximum RMSE reduction of 3.9\% in July. There is an increase of the RMSE occurring in February and April, of 0.1 \% and 0.2 \%, respectively, but the observed increase in RMSE for February and April is not reflected in the correlation coefficient ($\rho$) (Figure \ref{fig:statistics_CONUS}(b)), whose percentage change remains either equal to or greater than zero.

The monthly averaged change in correlation ($\rho$) shown in Figure \ref{fig:statistics_CONUS}(b) also illustrates an overall improvement across the year for PC altitudes near the FL. Such improvements occur mainly for June through September with a maximum increase in correlation of about 0.05 (11.5\% of original $\rho$ in July). Overall, results consistently show that the FL is the PC height around which the RMSE and the correlation are mostly improved. 

\subsection{FL-Based vs Cloud-top Height Parallax Correction Assessment}

\begin{figure}[h]
  \centering
  \includegraphics[width=0.3\textwidth]{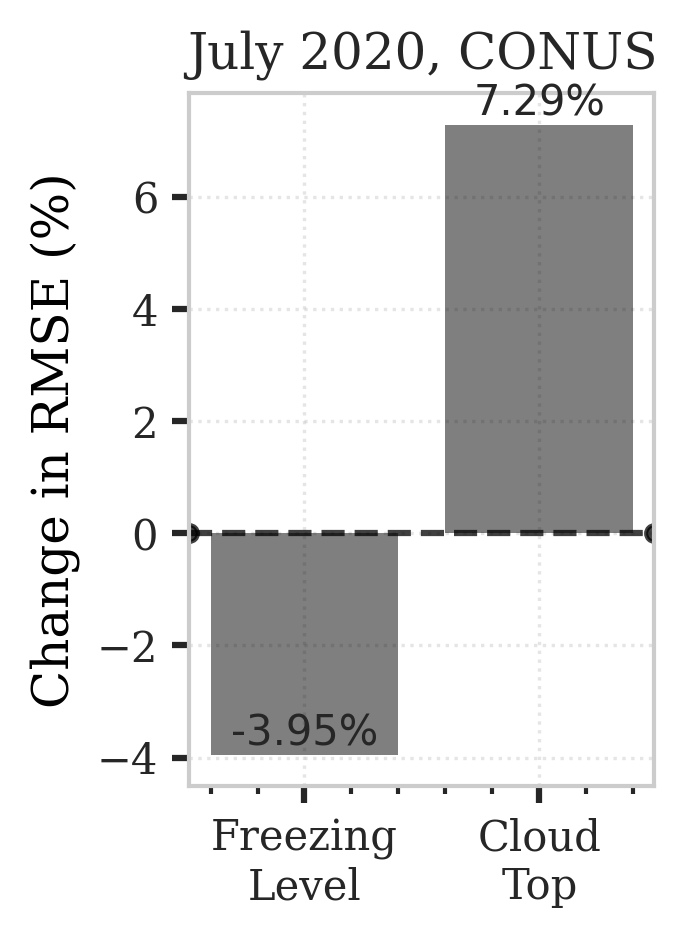}
  \caption{Change in precipitation RMSE of GMI Vs GV-MRMS over the CONUS during July 2020 for two methods of parallax correction; FL-based (3.95\% reduction) vs cloud-top-based (7.29\% increase).}
  \label{fig:FLvsCTop}
\end{figure}

A FL-based PC causes the most significant reduction (increase) in RMSE (correlation) on average over the CONUS, especially for July. To the best of the authors' knowledge, no other dynamic-height PC has been assessed specifically for PMW sensors. For comparison against another dynamic-height PC, a cloud-top-based PC is applied over the CONUS for July to asses its effects for the month with the biggest original RMSE (4 mm/h) and where the FL-based PC works better on average. Results of this comparison are shown in Figure \ref{fig:FLvsCTop}, where the effects of a cloud-top-based parallax correction increase the RMSE in 7.29\%.

\subsection{Precipitation-type Assessment of FL-Based PC}

\begin{figure}
  \centering
  \includegraphics[width=0.92\textwidth]{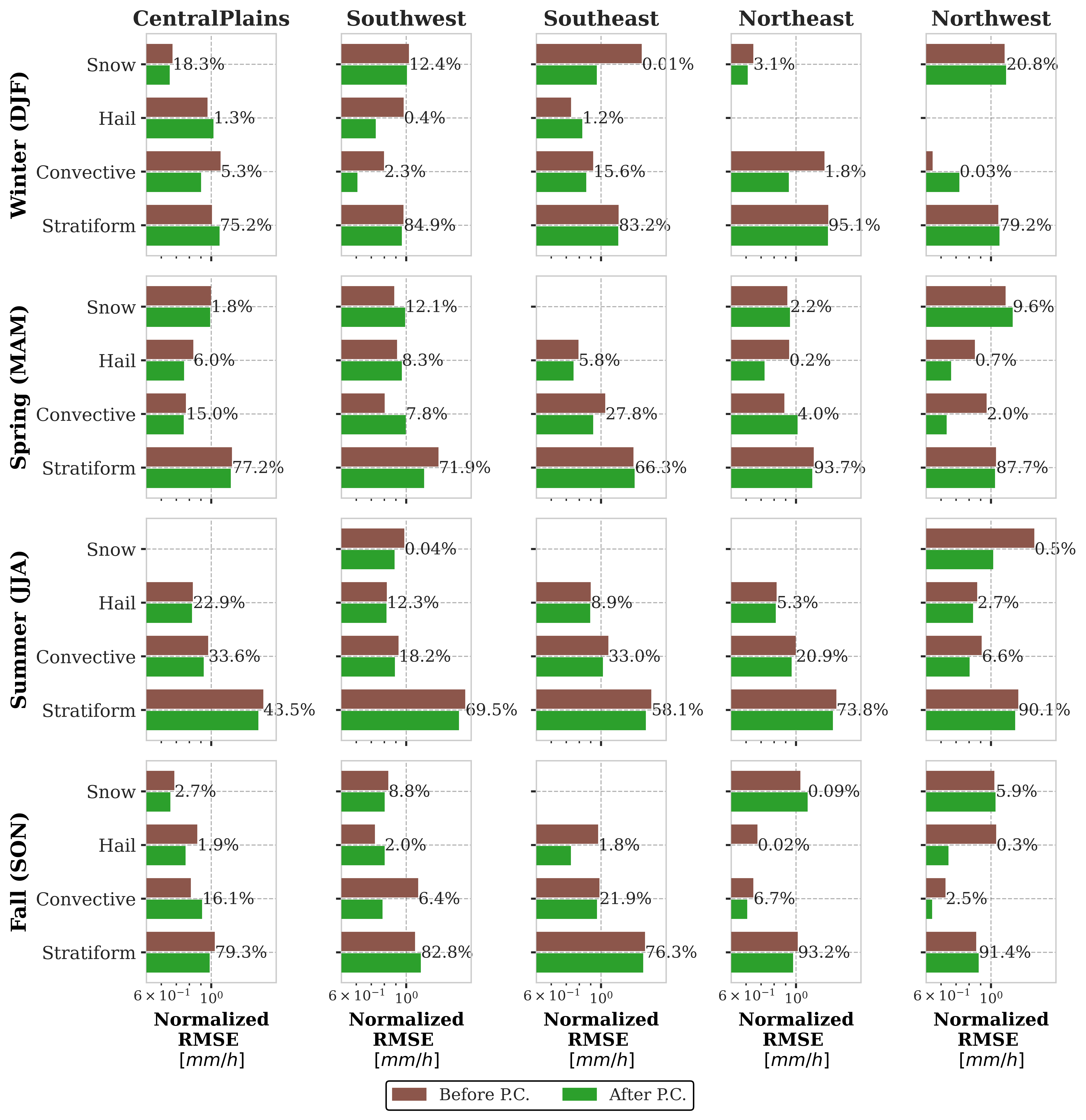}
  \caption{Normalized mean seasonal precipitation RMSE between uncorrected (brown bars; before PC) and parallax-corrected (at FL; green bars; after PC) GPROF retrievals per GV-MRMS reported precipitation types. The total contribution of every type is reported at the end of each bar as a percent of the total precipitation per region and season. For snow, hail, convective, and stratiform types, the selected pixels were those with intensity values greater than 0.2 mm/h either in GPROF or GV-MRMS. If no pixels of a certain type are found, the percentage is not shown.}
  \label{fig:RMSE_types}
\end{figure}

The purpose of this section is to understand what type of precipitation is most and least benefited from the use of a FL-based PC as proposed in this article. Precipitation types from GV-MRMS dataset, including snow, hail, convective and stratiform are examined. Figure \ref{fig:RMSE_types} illustrates the RMSE normalized by the mean GV-MRMS value of each precipitation type, within each of the studied CONUS region and season, between the original and parallax-corrected GPROF retrievals (at the FL) from October 1, 2019 to September 30, 2020.

Figure \ref{fig:RMSE_types} also shows the percentage of the total precipitation represented by each type by region and season. The stratiform type predominates across regions and seasons with more than 70\% of occurrence probability except in summer, where its lowest probability of occurrence is 42.9\%. During the summer, convective precipitation is the second most significant type across all regions. In contrast, snowfall is more prevalent during the other seasons than convective precipitation, particularly in the Southwest and Northwest.

The precipitation type with the highest errors relative to their magnitudes is the stratiform during summer, both before and after parallax correction. During summer, 0.5\% of precipitation is classified as snow in the Northwest, with the largest normalized RMSE, but it also substantially improves after PC. For other regions and seasons, the stratiform type consistently has the largest normalized RMSE compared to other types. However, in the Northwest, the snow type has a bigger normalized RMSE than the stratiform for every season.

From Figure \ref{fig:RMSE_types}, we observe that during summer the correction represents an improvement for every precipitation type for all regions. In every season, the stratiform events consistently show the slightest variations in RMSE before and after PC. During winter, the stratiform type shows no improvement after PC for any region. A different pattern is observed for hail and convection, with the RMSE showing a significant reduction during the fall and spring in the Northwest, and during the winter in the Southwest.

However, there are instances where the PC at the FL does not lead to an improvement in the RMSE: for convection in the winter in the Northwest, where convection is rare (accounting for only 0.03\% of total precipitation), as well as for convection in the fall in the Central Plains and in the spring in the Southwest.

\subsection{Individual Storm Assessment of Parallax-Corrected Retrievals}

\begin{figure}
  \centering
  \includegraphics[width=1.\textwidth]{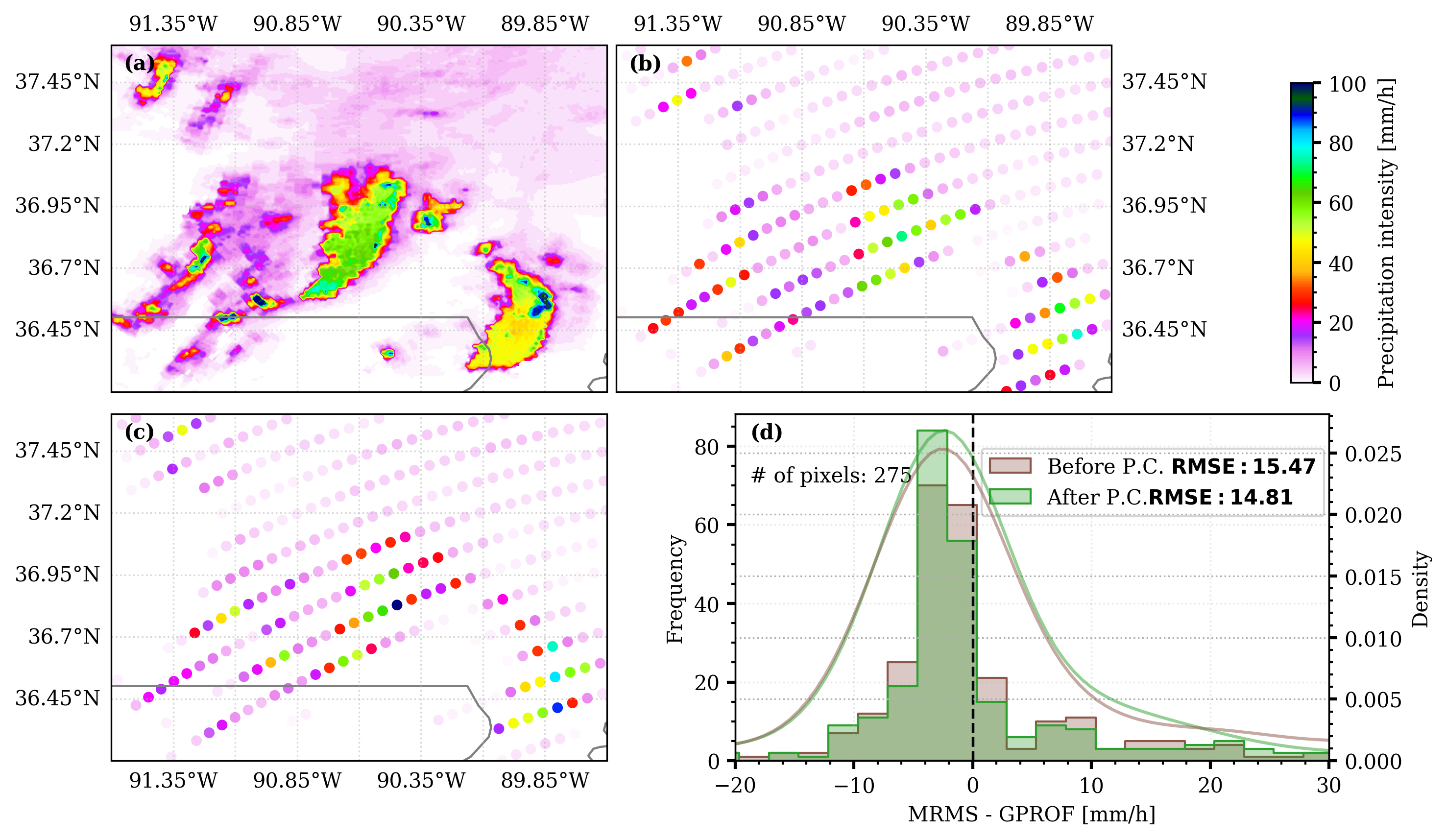}
  \caption{Instantaneous precipitation intensities for an event occurred in southeast Missouri on May 4, 2020 at 20:48 UTC. (a) GV-MRMS precipitation intensities (b) non-parallax-corrected precipitation at GMI pixels for the same storm and (c) parallax-corrected precipitation at GMI for the same storm (d) histograms and estimated probability density functions of the precipitation intensity differences between GV-MRMS and GMI for uncorrected (brown) and parallax-corrected (green) pixel positions.}
  \label{fig:event_improvement}
\end{figure}

This analysis was conducted to test the validity and usefulness of the proposed FL parallax-correcting strategy at the temporal scale of a singular storm event. Figure \ref{fig:event_improvement}(a) shows the GV-MRMS captured storm intensities for a precipitation event on May 4, 2020, in southeast Missouri. The storm covered approximately 20,000 square miles, with a significant storm core concentrated in about 1,000 square miles. (green region in the Figure \ref{fig:event_improvement} color bar). The storm had a maximum instantaneous intensity of about 90 mm/hr, according to GV-MRMS. 

The same event but looking at the matched pixels for uncorrected parallax GMI locations is shown in Figure \ref{fig:event_improvement} panel (b). Figure \ref{fig:event_improvement} panel (c) shows the GV-MRMS precipitation intensities but now plotted on collocated, parallax-corrected GMI pixels. The uncorrected GV-MRMS, panel (b) on \ref{fig:event_improvement}, shows a maximum intensity of around 80 mm/h on a single pixel, a pixel that is way below the maximum intensity shown on the corrected panel, i.e. \ref{fig:event_improvement} panel (c), that is nearest to 100 mm/h. The estimated probability distribution function (pdf) of the difference between GV-MRMS and GMI before and after the correction can be seen in Figure \ref{fig:event_improvement} panel (d). Overall, for the precipitation scene in  Figure \ref{fig:event_improvement} we observe lower RMSE from the parallax-corrected retrievals, but not a displacement of the errors pdf, estimated with the Kernel Density Estimate method \citep{Waskom2021}.
\newline

\begin{figure}[hbt!]
  \centering
  \includegraphics[width=0.9\textwidth]{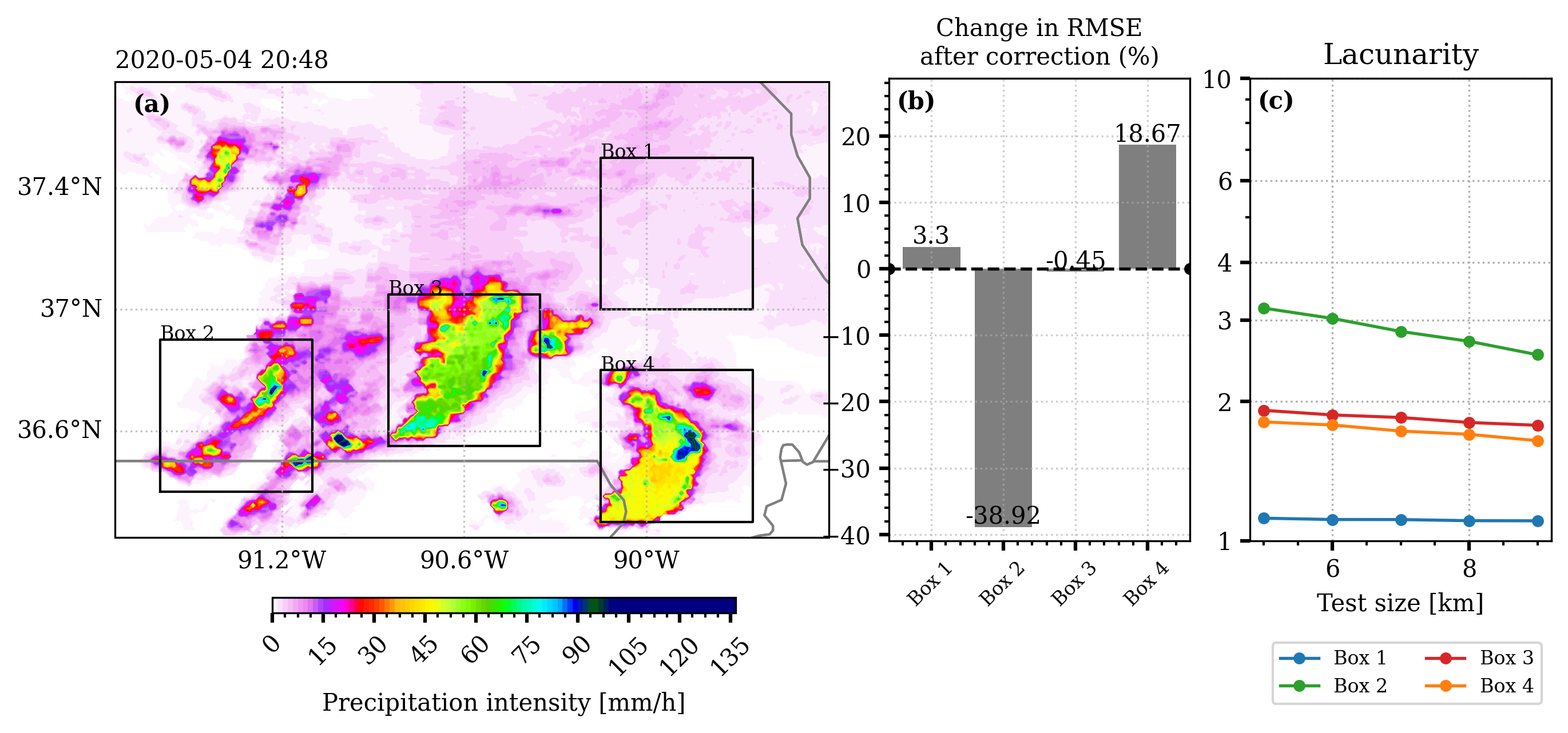}
  \caption{(a) GV-MRMS precipitation intensity field on May 4, 2020, at 20:48 UTC, with four boxes delineating spatial domains where the parallax correction at the FL was applied. (b) Percentage change in RMSE for each box in panel (a) following PC. (c) Lacunarity at multiple spatial scales for pixels within each box from panel (a).}
  \label{fig:event_1}
\end{figure}

\begin{figure}[hbt!]
  \centering
  \includegraphics[width=0.9\textwidth]{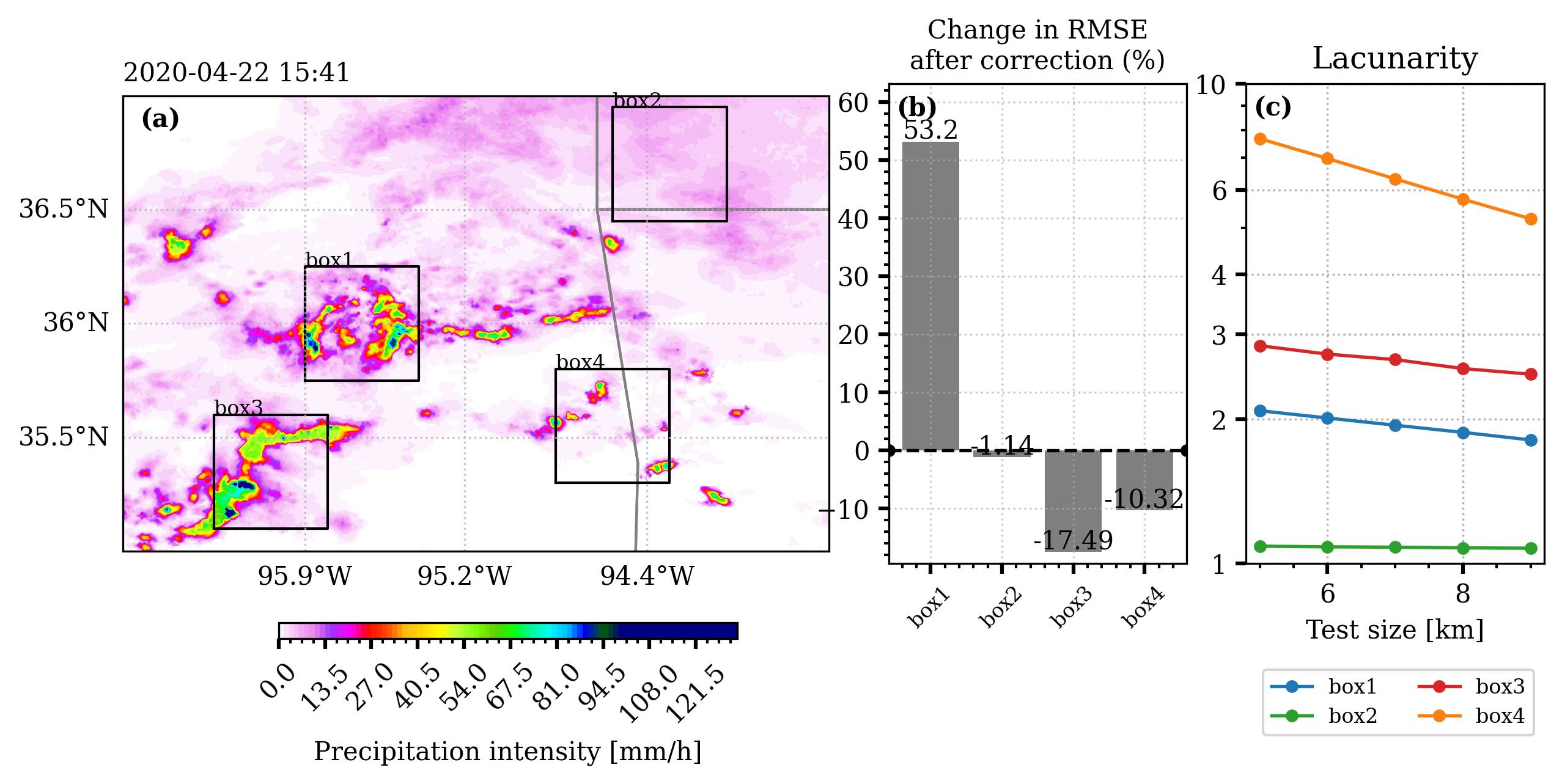}
  \caption{(a) GV-MRMS precipitation intensity field on April 22, 2020, at 15:41 UTC, with four boxes delineating spatial domains where the parallax correction at the FL was applied. (b) Percentage change in RMSE for each box in panel (a) following PC. (c) Lacunarity at multiple spatial scales for pixels within each box in panel (a).}
  \label{fig:event_2}
\end{figure}

\begin{figure}[hbt!]
  \centering
  \includegraphics[width=0.9\textwidth]{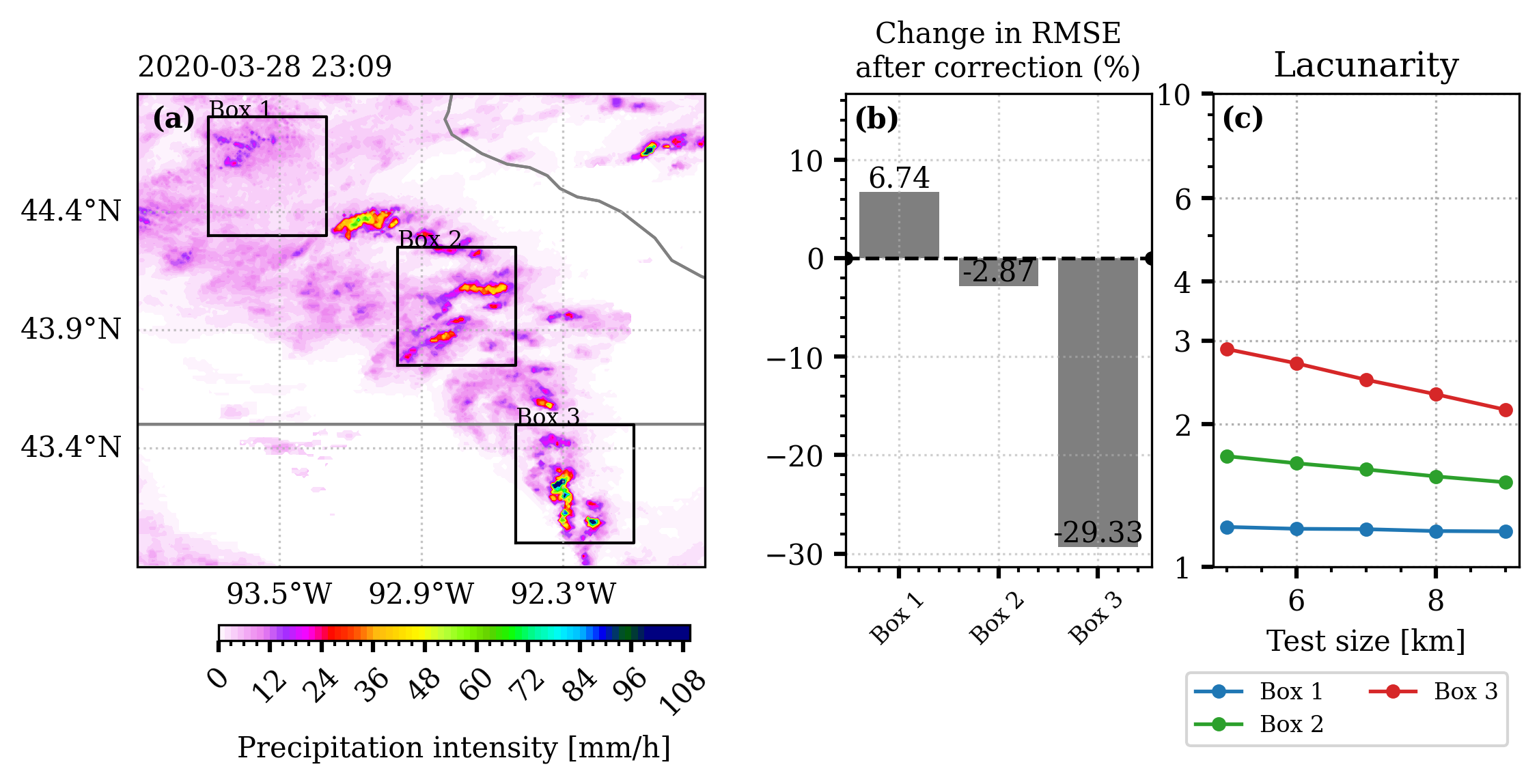}
  \caption{(a) GV-MRMS precipitation intensity field on March 28, 2020, at 23:09 UTC, with three boxes delineating spatial domains where the parallax correction at the FL was applied. (b) Percentage change in RMSE for each box in panel (a) following PC. (c) Lacunarity at multiple spatial scales for pixels within each box from panel (a).}
  \label{fig:event_3}
\end{figure}

To further assess the convenience of the FL-based PC, domains with different spatial patterns of precipitation within the same scene were selected, parallax-corrected, and their degree of spatial organization quantified. Panel (a) of  Figures \ref{fig:event_1}, \ref{fig:event_2}, and \ref{fig:event_3} show GV-MRMS scenes of different precipitation events captured both by GV-MRMS and GMI, and multiple sub-domain boxes of 0.5°x0.5° within each scene. For each sub-domain, panel (b) shows the percent change in RMSE after applying the PC to the GMI pixels inside. Panel (c) of each figure shows lacunarity values for multiple scales of analysis, from 5 km to 10 km (scales selected according to the theoretical GMI antenna gain function) \citep{guilloteau_multiscale_2020}.

In Figure \ref{fig:event_1}, we present a precipitation scene occurring on May 4, 2020 (same scene of Figure \ref{fig:event_improvement}). The sub-domain boxes inside contain multiple types of system behaviors: stratiform precipitation (Box 1), isolated intense nuclei of a few km of scale (Box 2), extensive and intense convective cells of tens of km of scale (Boxes 3 and 4). From the sub-domains of Figure \ref{fig:event_1} we found improvement only for Box 2, whose lacunarity value in panel (c) is above 2 for every analysis scale. The other sub-domains in Figure \ref{fig:event_1} did not reduce their RMSE after applying the PC, and their lacunarity value remains below 2 for every scale of analysis.

Another precipitation scene is presented in Figure \ref{fig:event_2} from April 22, 2020. The sub-domains include: intense precipitation clouds embedded in a matrix of lighter intensities (Box 1), a stratiform-like system (Box 2), intense but isolated nuclei surrounded by no-rain pixels (Box 3), very sparse and intense precipitation nuclei of a few km of scale (Box 4). Boxes 3 and 4 present significant reductions in RMSE after PC (17.5\% and 10.3\%, respectively), while Box 1 increases its RMSE by 53\%. Their calculated lacunarity values are above 2 for all analysis scales for boxes 3 and 4.

A final precipitation scene from March 28, 2020, is presented in Figure \ref{fig:event_3}. Its sub-domains include: a stratiform-like behavior (Box 1), some intense nuclei surrounded by light rain (Box 2), and intense and isolated precipitation nuclei (Box 3). Box 1 increased its RMSE by 6.7\%, while Boxes 2 and 3 reduced their RMSE after PC by 2.9\% and 29.3\%, respectively. The lacunary of Box 3 is above 2 for every scale, while boxes 1 and 2 lacunarities stay below 2.

%% file: 4_discussion.tex
 \section{Discussion}

This study illustrates the benefits of applying a physics-based, CONUS-statistically-optimized PC scheme for tilted satellite-orbiting PMW sensors, such as those aboard the GPM mission. We identified that a single-level correction at the water freezing temperature (calculated with ERA5 data) provides optimal results for nearly all cases analyzed. This novel finding is helpful from the operational perspective to reduce RMSE and increase correlation with GV-MRMS across diverse scenarios.

\begin{figure}
  \centering
  \includegraphics[width=1.\textwidth]{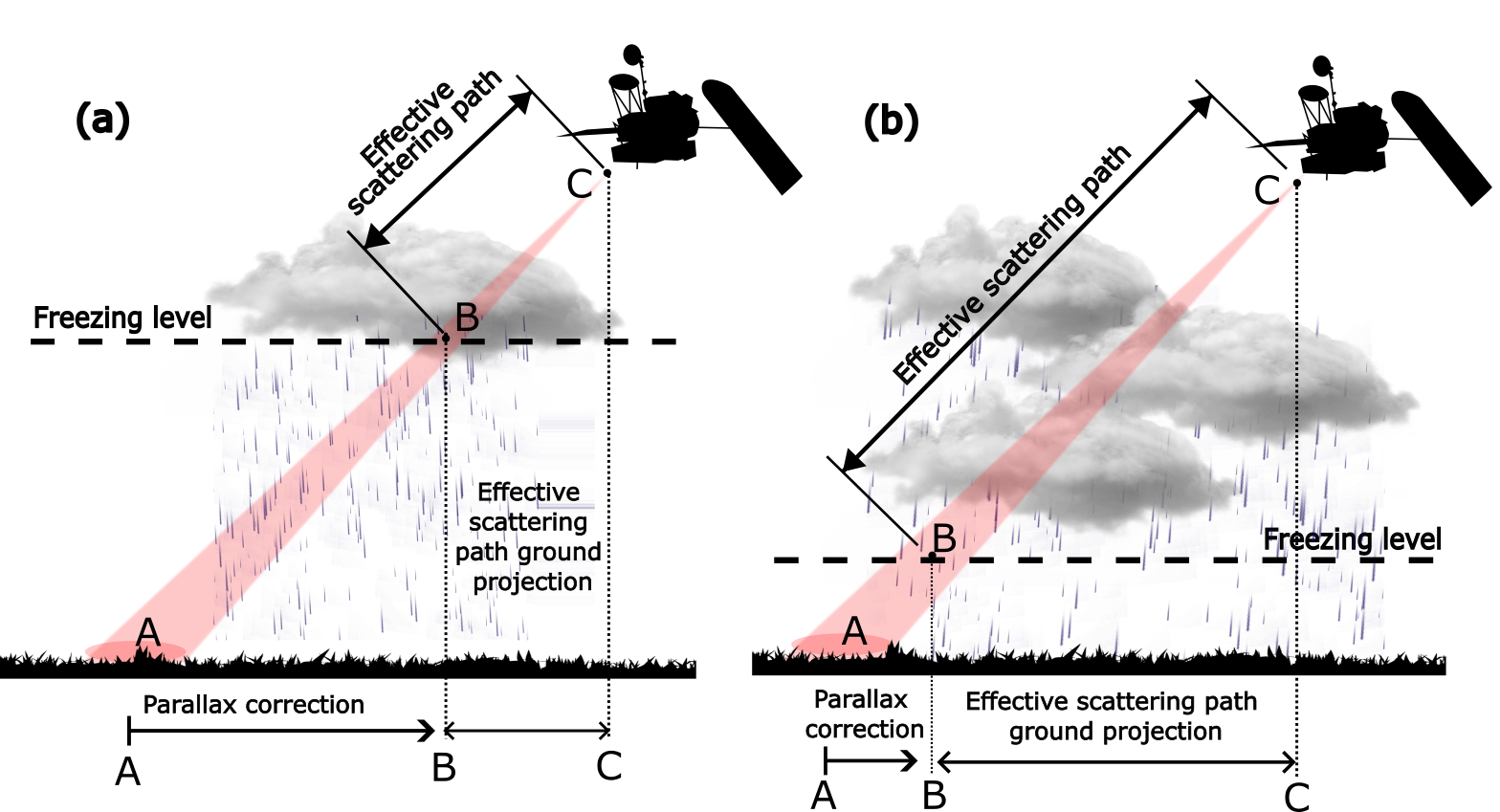}
  \caption{PC scheme using the freezing level as optimal PC height for (a) a relatively high freezing level, as in summer, and (b) a situation where the freezing level is relatively close to the surface as in winter. Vertical and horizontal scales are distorted for illustrative purposes.}
  \label{fig:parallax_vs_fl}
\end{figure}

The PMW sensor on board of GPM measures the decrease of upwelling radiation caused by ice hydrometeors normally found in the upper portions of convective clouds. It is not surprising that the height of PC, which causes the most significant improvement in the evaluation metrics, is close to the freezing level. Our research not only identified the optimal physically-based PC height for GMI, but also showcased the effectiveness of the developed framework in discovering a dynamic PC suitable for PMW sensors. Figure \ref{fig:parallax_vs_fl} illustrates some of the main take-aways from the results of this article.

When the freezing level is close to the surface (e.g., Figure \ref{fig:parallax_vs_fl} , panel (b)), as in winter in the northern hemisphere, the average improvement in evaluation metrics after applying the PC is not noticeable. The horizontal distance between region \textbf{A} and region \textbf{B}, is relatively short. In this case, the parallax-corrected coordinates fall in the same neighborhood. On the other hand, in cases where the freezing level is relatively high, as in summer (e.g., Figure \ref{fig:parallax_vs_fl}, panel (a)), the original \textbf{A} coordinates may not relate to the coordinates at \textbf{B} without the correction. 

In analyzing the performance of different precipitation types, we found that there is not a consistent improvement in performance after applying the correction. For instance, hail improves on average after the correction during spring and summer, but not during fall. It's important to note that snow and hail are not validated in the GV-MRMS dataset, which makes them more susceptible to the inherent biases of radar precipitation estimation, see Borga et al. (2002). It's also worth noting that the type of precipitation alone does not determine whether the parallax correction will improve the performance of the retrievals more than the season does. In summer, every type of precipitation shows a mean reduction in the normalized RMSE, with convective pixels showing a more consistent improvement.

\nocite{borga_longterm_2002}

On average, over CONUS, a reduction in RMSE of 1.6\% throughout the year and 4\% in July was achieved after the FL-based PC. Nevertheless, for the event-scale analysis, we detected cases where the proposed PC increased the RMSE of the scene by up to 53\% (as in Box 1, Figure \ref{fig:event_2}), as well as cases where the reduction in RMSE was of the order of 39\% (Box 2, \ref{fig:event_1}). By studying the spatial patterns of the boxes in Figures \ref{fig:event_1}, \ref{fig:event_2}, and \ref{fig:event_3} we observed a consistent result: for those scenes whose PC effect is a reduction in RMSE greater than 10\% we calculated relatively high lacunarities (above 2), whereas for the scenes that increased their RMSE after PC, their lacunarity values were relatively low compared to  scenes with the most important reductions in RMSE. 

To interpret and compare lacunarity values, it is important to recall that lacunarity is a dimensionless scale-dependent measure of the variation-to-mean ratio of mass within the test area \citep{plotnick_lacunarity_1996}. Therefore, if the total mass of pixels in the precipitation scene is evenly distributed across the test area, the variance and lacunarity will be low, as from Equation \ref{eqlac}. On the other hand, if the mass of the precipitation scene is concentrated in a few points, the variance will be high. Thus, the FL-based PC consistently improves the GMI event-scale representation of isolated intense nuclei with high lacunarity (calculated in a 0.5 x 0.5 degrees domain with a gliding-box test of a few kilometers).

One of the limitations of this study is the omission of surface type in the analysis. The precipitation estimation is based on the scattering of the emitted radiation, and the effect of different emissive backgrounds against the parallax mismatching has not yet been quantified. While it is not thought to be of high importance because higher frequency channels do not penetrate to the surface in cases of moderate rain, this was not specifically investigated. Another limitation is that the results are only applicable for GMI over land; over ocean, the best overall PC may not be the freezing level, given that the frequencies used to estimate the precipitation intensity over ocean rely more on the emission signals of rain itself. The best PC height may therefore be lower in the cloud than the ice scattering responsible for the parallax effect over land. We cannot assume that the improvement in the evaluation metrics for the FL-based PC from this study applies to other sensors. Independent assessment of each instrument's characteristics is necessary.

%% file: 5_conclussions.tex
\section{Conclusions}

This study used sets of GMI, GV-MRMS, and ERA-5 data over the CONUS, to (1) develop a framework to find the optimal PC height that improved precipitation retrievals of GMI in relation to GV-MRMS; and (2) evaluate the use of the FL-based PC algorithm and quantify the absolute reduction in precipitation estimation error. The main findings of this study are:

1. A statistically optimal atmospheric altitude for a parallax-correcting scheme that corresponds to the water freezing temperature height or freezing level. Applying this dynamic correction height hourly, according to ERA5 temperature profiles, will minimize systematic observed errors between precipitation intensities from GMI and GV-MRMS. We do not claim the FL to be the optimal PC height for every PMW instrument, but following this framework could help identify the PC height that improves performance for other sensors.

2. The warmest months in the northern hemisphere (June - September) see the most improvements in the precipitation retrievals with the biggest reduction in RMSE (i.e. GPROF Vs. GV-MRMS) of about 0.12 mm/h that explains a maximum of 4\% of CONUS average RMSE. For the coldest months, a slight to no improvement is observed. This is consistent with the smaller offsets in winter when the freezing level is near or at the ground.

3. There is not a consistent improvement or worsening of the retrievals after PC conditioned to a precipitation type (according to GV-MRMS pixel classification). For example, convection consistently improves after correction in winter, but not in spring. Over summer, all precipitation types of every regime show improvement, once more confirming the seasonal dependency of the proposed PC strategy.

4. At the event scale, the FL-based PC reduced the RMSE significantly (more than 10\%) in a consistent way for systems with specific spatial pattern characteristics; small and isolated precipitation nuclei scenes with a lacunarity value relatively high (above 2) calculated in a 0.5 x 0.5 degrees domain for the original GV-MRMS resolution between the 5 km and 10 km scale. We show examples of scenes like the one described in Figure \ref{fig:event_1} - Box 2, Figure \ref{fig:event_2} - Boxes 3 and 4, and in Figure \ref{fig:event_3} - Box 3.

5. The PC does not necessarily improve the ability of GPROF to represent extreme values. The original GMI retrievals tend to overestimate the low intensities and miss the highest intensities of an event. This characteristic still holds after applying the PC.